\documentclass[epj]{svjour}
\usepackage{graphics}

\begin{document}

\title{Extreme $\alpha$-clustering in the $^{18}$O nucleus.}

\author{E.D. Johnson\inst{1}   \and 
        G.V. Rogachev\inst{1}  \and
        V.Z. Goldberg\inst{2}  \and
        S. Brown\inst{1,3}     \and
        D. Robson\inst{1}      \and
        A.M. Crisp\inst{1}     \and
        P.D. Cottle\inst{1}    \and
        C. Fu\inst{2}          \and
        J. Giles\inst{1}       \and
        B.W. Green\inst{1}     \and
        K.W. Kemper\inst{1}    \and
        K. Lee\inst{1}         \and
        B.T. Roeder\inst{1,2}  \and
        R.E. Tribble\inst{2}
}
\institute{Department of Physics, Florida State University, Tallahassee, FL 32306 \and
           Cyclotron Institute, Texas A\&M University, College Station, TX 77843 \and
           The University of Surrey, Guildford, Surrey, UK
}

\date{Received: date / Revised version: date}

\abstract{
The structure of the $^{18}$O nucleus at excitation energies above
the $\alpha$ decay threshold was studied using $^{14}$C+$\alpha$
resonance elastic scattering. A number of states with large $\alpha$
reduced widths have been observed, indicating that the $\alpha$-cluster
degree of freedom plays an important role in this N$\ne$Z nucleus.
A 0$^{+}$ state with an $\alpha$ reduced width exceeding the single particle
limit was identified at an excitation energy of 9.9$\pm$0.3 MeV.
We discuss evidence that states of this kind are common in
light nuclei and give possible explanations of this feature.
}

\maketitle

After the discovery of the neutron it was understood that $\alpha$
particles cannot exist in the bulk of a nucleus at normal nuclear
density due to the need for antisymmetrization over all nucleons.
Still, $\alpha$ clustering manifests itself in light 4N nuclei such
as $^{8}$Be, $^{12}$C, $^{16}$O, and $^{20}$Ne through the existence
of twin quasi-rotational bands of states with alternating parities
and $\alpha$-particle reduced widths which are close
to the single particle limit (see recent review by M. Freer \cite{Freer07}
and references therein).

There are different approaches that attempt to describe the shell
model and the $\alpha$-cluster structure of nuclei on an equal
footing in order to shed light on the interplay between the $\alpha$-cluster
and single particle degrees of freedom \cite{Freer07}. Data on the
nucleon decay of $\alpha$-cluster states would be instrumental for
such efforts. However, this data is practically absent due to the
much higher nucleon decay thresholds in comparison with the thresholds
for the $\alpha$ decay in the 4N nuclei. It is more promising to
observe nucleon decay from the $\alpha$-cluster states in N\ensuremath{\neq}Z
nuclei where the nucleon and $\alpha$-particle thresholds are close
to each other. The study of non-self-conjugate nuclei has an advantage
in that one can investigate $\alpha$-cluster states in mirror systems
and use the Coulomb shift to extract information on the relationship
between the cluster and single particle degrees of freedom. Unfortunately,
the data on the $\alpha$-cluster structure of N\ensuremath{\neq}Z
nuclei are generally very limited.

Current interest in the $\alpha$-particle interaction with N\ensuremath{\neq}Z
nuclei is also strongly motivated by astrophysics \cite{Apra05}.
Even if astrophysical reactions involving helium do not proceed through
the strong $\alpha$-cluster states (because of high excitation energy),
these states can provide $\alpha$ width to the states that are closer
to the region of astrophysical interest through configuration mixing
\cite{Fu08}. However, the study is complicated by experimental difficulties
and the need for multi-channel analysis of many broad overlaping,
interfering resonances. This is illustrated by the previous investigations
of the $\alpha$+$^{14}$C interaction. In the first \cite{Morg70}
(and only) study of the resonances in the $\alpha$+$^{14}$C elastic
scattering, the experimental difficulties were minimized by using
a solid (radioactive) $^{14}$C target. However, only a few assignments
for some strong and sharp resonances were made. Goldberg et al. \cite{Gold04}
reanalyzed the data of \cite{Morg70} using a simplified version of
the \textbf{R}-matrix theory. In the analysis approach used in \cite{Gold04} 
only the elastic scattering channel was considered explicitly, while influence 
of all other decay channels was parametrized through a
$\Gamma_{\alpha}/\Gamma_{total}$ parameter for each resonance.  
They gave a few new assignments, but again many of them were tentative. 
The need for a more detailed investigation, better fit and new experimental data 
was pointed out in \cite{Gold04}. The $\beta$-delayed
$\alpha$ spectrum of $^{18}$N was measured in Refs. \cite{Zhao89,Buch07}.
Due to the high selectivity of $\beta$ decay only 1$^{-}$ states
were observed in the $^{18}$O spectrum. Several broad 1$^{-}$ states
were identified in \cite{Buch07}. In this
letter we report new measurements of $\alpha$+$^{14}$C resonance
elastic scattering made using the Thick Target Inverse Kinematics
(TTIK) method \cite{Arte90}. This experimental approach provided
good statistics and excitation functions which were remarkably
free of any background. The data provided a basis for a successful
analysis of the excitation functions using a complete multi-level,
multi-channel \textbf{R}-matrix approach \cite{lane58}. As a result,
we found that in $^{18}$O there is a system of strong $\alpha$-cluster
states. Both $\alpha$ and nucleon reduced widths of these $\alpha$-cluster 
states were determined from the multi-channel \textbf{R}-matrix fit. 
The main focus of our discussion in this letter is the surprising
finding of a state with an $\alpha$-particle width exceeding the
single particle limit. We interpret this as evidence for extreme 
$\alpha$-clustering and argue that existing experimental data on 
light 4N nuclei indicates that this is a common feature for nuclei 
in this mass range.

The experiment was carried out at the Florida State University John
D. Fox Superconducting Accelerator Laboratory. The 25 MeV $^{14}$C
beam was produced by an FN Tandem Van-de-Graff accelerator and directed
into the scattering chamber, which consisted of two compartments.
The first compartment was under vacuum, the second was filled with
99.9\% pure helium gas ($^{4}$He). The two compartments were separated
with a 1.27 $\mu m$ Havar foil. The intensity and quality
(focusing and alignment) of the beam was monitored using elastic scattering
off a gold target, placed in the middle of the vacuum compartment,
and from the Havar foil (energy of the beam is less than the C+Co 
Coulomb barrier). The helium gas pressure in the second compartment
was adjusted to stop the incoming beam before an array of Silicon
detectors located at a distance of $\sim$40 cm from the Havar foil.
Details of the method are given in \cite{EDJ-Thesis}.

The excitation functions for the $^{14}$C$+\alpha$ elastic scattering
covering the center-of-mass (c.m.) energy region of 2.0-4.5 MeV were
measured at 20 different angles. The observed background was less
than 1\%. Conversion of the laboratory
excitation functions into the c.m. was made on a bin-by-bin basis
using a computer code which takes into account the relevant experimental
conditions \cite{Rog99}. The accuracy of the absolute normalization
of the cross section, performed by the monitor detector using elastic
scattering of the $^{14}$C beam off the Havar foil, is 15\%. The
uncertainty in the specific energy loss of $^{14}$C in helium is
responsible for a 20 keV uncertainty in the absolute calibration of
the c.m. energy.

\begin{figure}

\resizebox{0.85\columnwidth}{!}{%
  \includegraphics{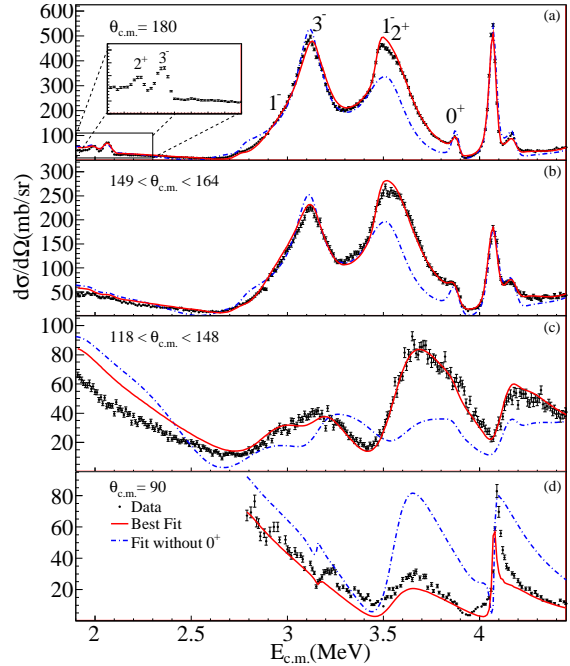}
}

\caption{\label{fig:spec0+} (Color online) The excitation functions for $\alpha$+$^{14}$C
elastic scattering measured at various angles, demonstrating the influence
of the broad J$^{\pi}=0^{+}$ state reported here. The top three panels
are data from the present experiment, while the bottom panel at 90$^{\circ}$
is taken from \cite{Morg70}. The experimental technique used here
yields a continuously varying scattering angle for each excitation
function except at 180$^{\circ}$; hence, the angular ranges shown
in the middle two panels overlap. The red curves show the best \textbf{R}-matrix
fit with the broad J$^{\pi}=0^{+}$ state at a resonance energy of
3.7 MeV (corresponding to an excitation energy of 9.9 MeV in $^{18}$O),
while the blue dash-dotted curve is the best fit without this state.
States with large $\alpha$ reduced width are labeled.}

\end{figure}

The analysis of the excitation functions was performed using a multi-level,
multi-channel \textbf{R}-matrix approach \cite{lane58}. The $^{14}$C+$\alpha$
excitation functions measured in this experiment are continuous which excludes
the possibility of missing a narrow resonance. The sensitivity of the data 
is demonstrated by the fact that even the 2$^+$ state at 8.213 MeV 
(1.986 MeV c.m.), which has width of ~1 keV, is still clearly visible in the 
180$^\circ$ excitation function (see inset in Fig. \ref{fig:spec0+}). 
In addition, data from previous measurements of the $^{14}$C($\alpha,$$\alpha$) 
\cite{Morg70} and $^{14}$C($\alpha$,n) \cite{Bair66} excitation functions were 
used. The overall fit was very good with $\chi^{2}/\nu$=1.64 for the c.m.\ 
energy range 2.65-4.45 MeV. The \textbf{R}-matrix fit to the 
$^{14}$C($\alpha$,$\alpha$) data is shown in Fig. \ref{fig:spec0+}. As can be 
expected most of the states also have substantial neutron widths, which are 
obtained through the ($\alpha$,$\alpha$) fit.  We verified that the 
$^{14}$C($\alpha$,n) total cross section from Ref. \cite{Bair66} is reproduced 
rather well by \textbf{R}-matrix calculations performed using the parameters 
from the $^{14}$C($\alpha$,$\alpha$) fit. One can notice that the 
\textbf{R}-matrix fit underestimates the experimental cross section at the 
lowest energies for angles far from 180$^{\circ}$. This is understood to be an 
effect of the finite dimensions of the beam spot.  

The fourth panel in Fig. \ref{fig:spec0+} is 
90$^{\circ}$ data from \cite{Morg70}. The spectrum at 90$^{\circ}$ in c.m. is only 
influenced by states with even angular momentum and positive parity. This is an 
important simplification and makes the spectrum at 90$^{\circ}$ very valuable for 
the \textbf{R}-matrix analysis. Clearly, our data contains this information. 
However, for the purpose of a more clear representation of the data we used the 
spectrum of \cite{Morg70} at 90$^{\circ}$ rather than construct the 90$^{\circ}$ 
spectrum from several different detectors.

Twenty-four resonances were used to 
fit the data, some of them were previously known. Detailed discussion of 
the analysis procedure and notes on the properties of each 
state will be given in the follow up paper. Levels with large 
$\alpha$-cluster reduced widths 
($\theta_{\alpha}^{2}>0.1$) are given in Table \ref{tab:levels}.

Five levels with dimensionless $\alpha$ reduced width greater than $0.1$ have 
been observed in $^{18}$O in the narrow excitation energy range between 9.1 
and 9.9 MeV. Three of them have been suggested in previous publications 
\cite{Zhao89,Gold04,Buch07}. The strong $\alpha$-cluster state at $9.0\pm0.2$ 
MeV was first suggested in \cite{Zhao89}, where the 1$^-$ spin parity 
assignment was made on the basis of population of this state in $^{18}$N $\beta$ 
decay. A more recent $^{18}$N $\beta$ decay experiment \cite{Buch07} confirmed the 
1$^-$ state at 9.16 MeV with a width of $\Gamma=420\pm200$ keV. Our 
\textbf{R}-matrix fit requires a 1$^-$ state at an excitation energy of 
9.17$\pm$0.03 MeV in good agreement with \cite{Buch07}. The width of this state 
is lower in the present work (230$\pm$50 keV) but still within the error bars of 
\cite{Buch07}. Another 1$^-$ state observed at 9.85$\pm$0.5 MeV in \cite{Buch07} 
with a width of 560$\pm$200 keV is in very good agreement with the results of 
this work. The 3$^-$ at 9.39 MeV was previously suggested in \cite{Gold04} and 
has excellent agreement with the present results. The strong $\alpha$-cluster 
2$^+$ state at 9.77 MeV is identified for the first time in this work.

\begin{figure}

\resizebox{0.85\columnwidth}{!}{%
  \includegraphics{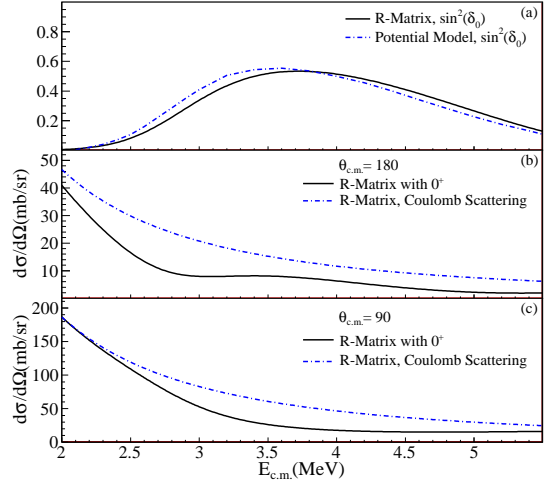}
}

\caption{\label{fig:phase_shift} 
The top panel is the $\sin^2\delta_{\ell=0}$ from the \textbf{R}-matrix fit (solid curve)
compared to the potential model prediction (dash-dotted curve). The middle and the 
bottom panel demonstrate influence of the broad 0$^+$ state on the cross section at 
180$^{\circ}$ and 90$^{\circ}$ c.m. respectively. The dash-dotted curve corresponds to 
the Coulomb scattering cross section and the solid curve shows the differential
cross section from the interference of the broad 0$^+$ with Coulomb scattering.}
\end{figure}

The most surprising finding of this work was the observation
of a very broad 0$^{+}$ state at 9.90$\pm$0.3 MeV. 
Generally it is not easy to identify very broad resonances because they are
disguised by interference with the sharp ones. This is especially
true for the 0$^{+}$ resonances. Nevertheless, existence of the 
broad $\alpha$-cluster 0$^{+}$ state in the spectrum of $^{18}$O is
certain. The effect of this state on the cross section
is demonstrated in Fig. \ref{fig:phase_shift}. The cross section is lowered 
dramatically due to the destructive interference of the 0$^+$ state with Coulomb 
scattering. The Rutherford cross section and the Rutherford with the broad 0$^+$ 
state at 9.90 MeV (3.7 MeV c.m.) are shown 
at two c.m. angles, 90$^{\circ}$ and 180$^{\circ}$, in Fig. \ref{fig:phase_shift}.  
Without this destructive interference it is not possible to reproduce
the experimental data. As seen in the bottom panel of Fig. \ref{fig:spec0+}, 
the cross section calculated without the broad 0$^{+}$ state is 
significantly larger than the experimental one. This is a clear 
indication of a 0$^{+}$ resonance. All other resonances would not 
produce the right interference with the Rutherford scattering and 
could not be broad enough. One
can follow the effect of the 0$^{+}$ level at different angles in
Fig. \ref{fig:spec0+} and reach the same conclusion. We also considered the 
possibility of two narrower 0$^{+}$ states instead of one. The decisive factor 
against it was that the characteristic interference pattern between two nearby 
0$^{+}$ states was not observed experimentally.
The $\alpha$ particle
reduced width amplitude of the 0$^{+}$ state with channel
radii of 5.2 and 6.5 fm are 1.38 and 0.66 MeV$^{1/2}$, respectively.
Formally, both values exceed the single-particle limit. Using a classical approach
one can interpret this as evidence that the $\alpha$ particle resides
at a large distance from the $^{14}$C core.

\begin{table}
\caption{ \label{tab:levels}
Levels with large $\alpha$-reduced width in $^{18}$O.
$\Gamma_{tot}$, $\Gamma_{\alpha}$ and $\Gamma_{n}$ are the total
and partial $\alpha$ and neutron widths, respectively. 
$\theta_{\alpha}^{2}=\gamma_{\alpha}^{2}/\gamma_{SP}^{2}$
is the dimensionless reduced width for the $\alpha$ channel, where
$\gamma_{SP}^{2}=\hbar^{2}/\mu R^{2}$ is the single particle limit,
calculated at channel radius 5.2 fm.}
\begin{tabular}{llllll}
\hline\noalign{\smallskip} 
$E_{exc}$  & $J^{\pi}$  & $\Gamma_{tot}$  & $\Gamma_{\alpha}$  & $\theta_{\alpha}^{2}$  & $\Gamma_{n}$  \\
(MeV)  &  & (keV)  & (keV)  &  & (keV)  \\
\noalign{\smallskip}\hline\noalign{\smallskip}
9.17(3)       & $1^{-}$  & 229(50)        & 200(50)    & 0.24  & 29(7)         \\
9.39(3)       & $3^{-}$  & 151(50)        & 100(48)    & 0.45  & 51(15)        \\
9.75(4)       & $1^{-}$  & 628(100)       & 585(100)   & 0.43  & 43(10)        \\
9.77(3)       & $2^{+}$  & 251(50)        & 172(45)    & 0.20  & 79(20)        \\
9.9(3)*       & $0^{+}$  & 2100(500)**    & 2100(500)  & 2.6   &  N/S***       \\      
\noalign{\smallskip}\hline
\end{tabular}
\\
Error bars are dominated by systematic errors in absolute energy calibration and absolute 
cross section normalization.\\ 
$*$ Excitation energy of the 0$^+$ state is defined as maximum of the $\sin^2\delta_\circ$,
where $\delta_\circ$ is the $\ell$=0 phase shift (see Fig. \ref{fig:phase_shift})\\
$**$ Width of the 0$^+$ state is defined as FWHM of the $\sin^2\delta_\circ$.\\
$***$ The R-matrix fit is not sensitive to the neutron partial width of the 0$^+$ 
state within reasonable limits.
\end{table}

A more detailed description of the observed 0$^{+}$ state in $^{18}$O
can be given using the potential model approach. The parameters of
this model were extracted starting from the potential model for $^{8}$Be
given in \cite{Buck84}, which provides an accurate description of
the s-wave $\alpha$-$\alpha$ phase shift over a large energy range.
First, this potential model was used to investigate the $^{16}$O+$\alpha$
interaction by assuming the $\alpha$-cluster model \cite{Wheeler37}
for the ground state of $^{16}$O, properly modified \cite{Robson79}
to include antisymmetry and the strong repulsion between nucleons.
The strong interaction between the incoming $\alpha$ and the target
``alphas'' is obtained by folding the Buck interaction \cite{Buck84}
with the target $\alpha$-density in the ground state and produces
a nuclear potential with several Pauli forbidden states. These are
removed by adding a repulsive Gaussian potential in the relative $\alpha$-$^{16}$O
radial coordinate. The result is to produce a bound state for the
ground state of $^{20}$Ne and a broad $0^{+}$ state at $4$ MeV in
c.m.\ as observed in \cite{McDer60} with the same $0^{+}$ phase
shift behavior. The corresponding density distribution for this $0^{+}$
resonance has an inner peak at 2 fm and an outer peak at 5 fm. The
outer peak is at a separation distance beyond the sum of the charge
radii of $^{4}$He and $^{16}$O and it contains most of the probability
indicating that the broad $0^{+}$ resonance in $^{20}$Ne appears
to be a state with extreme $\alpha$-clustering.

The $^{18}$O system was then investigated using the $^{20}$Ne potential
as a starting point but small changes in the strengths of the folded
and repulsive potentials were used to reproduce the
ground state of $^{18}$O. This potential produces a broad $0^{+}$ state 
at 3.5 MeV in c.m. The experimental s-wave resonance phase shift from
the \textbf{R}-matrix fit (solid black curve in 
Fig. \ref{fig:phase_shift}a) is reproduced rather well by such a 
potential (dash-dotted blue curve in 
Fig. \ref{fig:phase_shift}a). The density
distribution for the broad $0^{+}$ state in $^{18}$O is similar
to the one in $^{20}$Ne. The outer peak is at 5.5 fm which is considerably
larger than the sum of the charge radii for $^{4}$He and $^{14}$C
and even larger than the $^{20}$Ne outer peak radius. This latter
result is presumably due to the larger Coulomb barrier in the $^{20}$Ne
system. Both systems appear to show well separated $\alpha$-cluster
configurations that correspond to extreme $\alpha$-clustering. 

Due to the large distance between the $\alpha$ cluster and the core,
one can speculate that levels of this kind should be a general feature
of the interaction between an $\alpha$ particle and a core nucleus
which is independent of the specific structure of the core. If this
is the case then resonances of this kind should be present in all
nearby nuclei at excitation energies on the order of a few MeV above
the $\alpha$ decay threshold. Indeed, a broad (3 MeV) $0^{+}$ level
at an excitation energy of 10.3 MeV ($\sim$3 MeV above the $\alpha$
threshold) appears in $^{12}$C \cite{Till98}. Recently, the parameters
of this $0^{+}$ level were revised in measurements of the 3$\alpha$
decay of this level populated after $\beta$ decay of $^{12}$B or
$^{12}$N \cite{Fyn05,Dig05}. The new parameters, as given in \cite{Dig05},
are an excitation energy of 10.73 MeV and a width of 1.72 MeV. However,
there are still significant uncertainties in these values. 

There is much more controversy related to a possible broad $0^{+}$
level in $^{16}$O at $\sim$11.3 MeV excitation energy ($\sim$4
MeV above the threshold) \cite{ajze84}, since there are conflicting
results for this state \cite{Bitt54,Bohne69,Ful69,Clark68,till93}.
However, the most recent article presented clear evidence for broad
$0^{+}$ strength at the excitation energy in question \cite{Lui01,Lui-Priv}.
Finally, a broad $0^{+}$ level ($\Gamma>800$ keV) at $\sim$4
MeV above the $\alpha$ decay threshold, was observed in $^{20}$Ne
\cite{McDer60}. This result has not been questioned since it was
first reported.

One may refer to these states as ``$\alpha$-halo'' states. This term was first 
suggested in \cite{Funaki05} and has merit. Also, one should not link directly 
unbound $\alpha$-halo resonances with bound neutron halo configurations found
in some weakly bound neutron rich nuclei.

While the purely single $\alpha$ particle nature of these states is clear 
it is interesting to consider the possible quantum mechanical reasons behind 
the emergence of such pure configurations. The answer may be a ``super-radiance" 
phenomenon (see for example \cite{Volya03}). Super-radiance emerges in the limit 
of strong coupling to the continuum (broad strongly overlaping resonances of the 
same spin-parity) and represents accumulation of the total summed $\alpha$ width 
from the unperturbed intrinsic states by one very broad (super-radiant) resonance. 
Since the centrifugal barrier is absent for $\ell=0$ states the condition of 
strong continuum coupling is more likely realized for the 0$^+$ states than for 
the states of any other spin-parity.      

In summary, we are only in the beginning phases of studying the $\alpha$-cluster
structure of light N$\neq$Z nuclei. Here we report a measurement
of the $^{14}$C+$\alpha$ elastic scattering excitation functions
including a successful analysis using the complete \textbf{R}-matrix
framework. We identified that the $\alpha$-cluster states in $^{18}$O
have many surprising properties, foremost of which is the discovery
of a broad, $\ell$=0 state which we suggest may be present in
other nuclei in this mass range. This conclusion is strengthened through comparison
with previous results \cite{McDer60,Till98,Fyn05,Dig05,ajze84,Lui01}.
There is much work still ahead to prove this point. The discovery
of the broad states in question within odd-even nuclei is an especially
important and difficult task. Nevertheless, the possible influence
that these broad low spin resonances may have on astrophysically important
reaction rates along with the insight they can give into $\alpha$
clustering justifies these efforts.

The authors are grateful to Prof. A. Volya for useful comments 
and discussions. This work was supported by the NSF under grant number
PHY-456463 and the DOE under grant number DE-FG02-93ER40773. 

\bibliographystyle{epj}
\bibliography{myrefs_ver3_short}

\end{document}